\documentstyle[prc,preprint,aps,epsfig]{revtex}
\tightenlines

\newcommand{\veps}{\mbox{\boldmath $\epsilon$ \unboldmath}}

\begin{document}

\title{\bf Contributions of vector meson photoproduction
to the Gerasimov-Drell-Hearn sum rule}

\author{Qiang Zhao$^1$\thanks{Electronic address: 
qiang.zhao@surrey.ac.uk.}, J.S. Al-Khalili$^1$, and C. Bennhold$^2$}
\address{1) Department of Physics, University of Surrey, Guildford, Surrey 
GU2 7XH, United Kingdom}
\address{2) Department of Physics, Center for Nuclear Studies,\\
 The George Washington University, Washington, D.C. 20052}

%\date{\today}

\maketitle  
  
\begin{abstract}

An improved version of 
a recently developed quark model approach 
to vector meson photoproduction is applied to the investigation
of contributions of vector meson photoproduction
to the Gerasimov-Drell-Hearn (GDH) sum rule. 
We find that the sum rule converges at a few GeV's.
Contributions to the proton channel are found to be
small while to the neutron are relatively large.

\end{abstract}
\vskip 0.5cm


\vskip 0.5cm
%\newpage

%\section{Introduction}
The GDH sum rule~\cite{gdh} connects the nucleon
resonance phenomena to the nucleon's magnetic moments, which 
are static properties of the groundstate nucleons,
\begin{equation}\label{gdh-01}
I_{GDH}=\int^\infty_{\nu_0} 
[\sigma_{1/2}(\nu)-\sigma_{3/2}(\nu)]
\frac{d\nu}{\nu}=-\frac{2\pi^2\alpha_e\kappa^2}{m^2_N},
\end{equation}
where $m_N$ is the nucleon mass; $\alpha_e$ is the fine structure 
constant; $\kappa$ is the nucleon anomalous magnetic moment;
$\sigma_{1/2}$ and $\sigma_{3/2}$ respectively 
denote photoabsorption cross sections for the nucleon and photon
helicities anti-parallel and parallel with each other;
$\nu_0$ denotes the threshold energy for single pion production in lab system.
Experimental and theoretical studies of exclusive reactions provide another
means of testing this sum rule, which can lead to 
better understanding of the internal degrees of freedom
of nucleons.
The availability of high intensive electron and photon beam
facilities gives access to precise measurements of meson photoproduction.
Recently, the GDH- and A2- Collaboration reported their 
results at the photon energies from 200 to 800 MeV~\cite{gdh-a2}.
Other experimental projects 
at JLab, MAMI, GRAAL, ELSA, and SPring8 
will make it possible to test this sum rule independently and
extend it to higher energies. 
In theory, extensive investigations
~\cite{karliner-73,workman-92,burkert-93,arndt-96,hanstein-98,drechsel-1,drechsel-2,tiator} 
from the low-energy limit to around 2 GeV photon 
energies have been carried out for the light
meson production channels. The most recent study 
by the Mainz group~\cite{tiator} 
including $\pi N$, $\eta N$, $\pi\pi B$ (Born terms), 
and $\pi\pi D_{13}(1535)$
showed that contributions from those channels
accounted for the sum rule for the proton up to 97\%, while 
large descrepancies were found for the neutron. 
Interestingly, these theoretical evaluations 
including the above channels led to results 
smaller than the absolute values for both proton and neutron,
which suggests that contributions from higher
energy production channels might add instead of 
cancel certain terms in the exclusive sum $I_{GDH}$ in Eq.~(\ref{gdh-01}).
In Ref.~\cite{sumowidagdo} contributions of the kaon photoproduction
to the GDH sum rule was estimated up to 2 GeV. 
It was found that the kaon contribution increased the
calculated value of GDH sum rule for the neutron, while might decrease
that for the proton. 
To satisfy the GDH sum rule by exclusive 
study, one might have to go up to a higher energy region.
On the one hand, there might be still significant contributions
from those light meson productions. On the other hand,
open channels above the kaon channels could start to play a role. 
This is one of our motivations for 
investigating the contributions from the
vector meson production channels,
for which the thresholds are just above the kaon 
production. Kinematically, 
the light vector meson ($\omega$ and $\rho$) production
is the next contribution that should be included in the 
exclusive calculations.
Certainly, 
the energy suppression in the integral
could have made the vector meson contribution trivial.
But such an effort is by no means trivial due to 
the diffractive 
features in vector meson photoproduction.
It would be a challenge for any model to show how the spin-dependent
feature changes to spin-independent at high energies
as required by the sum rule.
A reasonable estimation of the contributions of vector meson 
photoproduction to the GDH sum rule will also
provide a test for the model.

It should be noted that
in Ref.~\cite{bianchi-99} contributions above the resonance
regions to the GDH sum rule 
were estimated using a Regge parametrization, which 
accounted for a large fraction of discrepancies between the sum rule
and contributions from the single pion photoproduction.
This result could suggest that an explicit study of higher threshold 
processes should be necessary.

We employ a recently developed
quark model approach 
to vector meson photoproduction
~\cite{zhao-plb98,zhao-prc98,zhao-npa99,zhao-omega,zhao-plb01}
for this purpose.
It allows not only the study of resonance excitations 
in the $\omega$, $\rho$ and $\phi$ meson photoproduction, 
but also the inclusion of the diffractive contributions
with a mixed Pomeron exchange model. 
Thus, the model can be extended to photon energies of a few GeV's.
Unlike the study of various polarization observables, 
the most essential question 
arising from the sum rule study is the role played 
by the diffractive processes. 
In principle, the sum rule 
requires integrating the cross section difference 
up to infinite energy, while in vector meson 
production the diffractive cross sections 
tend to a constant value at high energies. 
This leaves the question of whether the spin-dependent terms 
can be averaged out efficiently at high energies. 
An analytical illustration
is not available yet. However, qualitatively, based on this 
specific model, we can see later that this approach 
allows us to make an energy cut at a few GeV's, where 
the spin-dependent terms have been sufficiently suppressed and averaged, 
and the dominant term in the diffractive process has spin-independent
feature. 
As a consequence, the integral of 
the cross section difference
will converge and be reliably estimated at a few GeV's.

The transition amplitude for $\gamma N \to V N^\prime$ 
can be expressed by 12 independent helicity amplitudes: 
\begin{eqnarray}
H_{1\lambda_v}=\langle \lambda_v, \lambda_f=+1/2 |{\cal T} 
|\lambda_\gamma=+1, \lambda_i=-1/2 \rangle, \nonumber\\
H_{2\lambda_v}=\langle \lambda_v, \lambda_f=+1/2 |{\cal T} 
|\lambda_\gamma=+1, \lambda_i=+1/2 \rangle, \nonumber\\
H_{3\lambda_v}=\langle \lambda_v, \lambda_f=-1/2 |{\cal T} 
|\lambda_\gamma=+1, \lambda_i=-1/2 \rangle, \nonumber\\
H_{4\lambda_v}=\langle \lambda_v, \lambda_f=-1/2 |{\cal T} 
|\lambda_\gamma=+1, \lambda_i=+1/2 \rangle, 
\end{eqnarray}
where ${\cal T}$ is the dynamic operator for this transition;
$\lambda_\gamma$ and $\lambda_v$ ($=0,\ \pm 1$) are helicities for the 
incident photon and outgoing vector meson, respectively,
while $\lambda_i$ and $\lambda_f$ are helicities for the 
initial and final state nucleons.
For simplicity, we have fixed the photon polarization vector 
as $\veps_\gamma=-(1, i, 0)/\sqrt{2}$. Parity conservation
will give amplitudes for $\veps_\gamma=(1, -i, 0)/\sqrt{2}$,
which are not independent from the former. 
For exclusive photoproduction, apart from the kinematic 
factors, one can see that
\begin{eqnarray} 
\sigma^v_{1/2}&\propto & H^2_{1\lambda_v}+H^2_{3\lambda_v},\nonumber\\
\sigma^v_{3/2}&\propto & H^2_{2\lambda_v}+H^2_{4\lambda_v}. 
\end{eqnarray}
The contribution from vector meson photoproduction
is expressed as 
\begin{equation}
I^v_{GDH}=\int^\infty_{\nu_v} 
\Delta\sigma(\nu)\frac{d\nu}{\nu}
=-\frac{2\pi^2\alpha_e \kappa^2_v}{m^2_N},
\end{equation}
where $\Delta\sigma(\nu)\equiv\sigma^v_{1/2}-\sigma^v_{3/2}$ 
is the photoabsorption cross section difference and the total cross section
for the unpolarized vector meson production
is $\sigma_{tot}=(\sigma^v_{1/2}+\sigma^v_{3/2})/2$; $\kappa_v$ 
denotes the contribution from vector meson production to the nucleon
anomalous magnetic moment;
$\nu_v$ is the threshold energy for the vector meson 
in the lab system (i.e., the nucleon rest frame). 
For the $\omega$ meson, $\nu_\omega=1.108$ GeV, while for 
$\rho$, $\nu_\rho=1.086$ GeV. 
Using the recent quark model developed by Zhao {\it et al}.
~\cite{zhao-plb98,zhao-prc98,zhao-npa99,zhao-omega,zhao-plb01},
we shall explicitly calculate the 
photoabsorption cross sections $\sigma^v_{1/2}$ and $\sigma^v_{3/2}$.

In this model, there are
three processes that contribute to the transition 
amplitudes for the neutral vector meson photoproduction:
(i) the {\it s}- and {\it u}-channel resonance excitations
and the nucleon pole terms; (ii) the {\it t}-channel 
light meson exchange (i.e., pion exchange in $\omega$ production
and $\sigma$ meson exchange in $\rho$ production);
and (iii) the {\it t}-channel Pomeron exchange in the neutral 
vector meson production (i.e., $\gamma N\to \omega N$ and
$\gamma N \to \rho^0 N$).

In the helicity frame, the {\it s}-channel resonance excitation
amplitude can be expressed as the product of the resonance 
electromagnetic excitation helicity amplitude $A^\gamma_{\Lambda_i}$
($\Lambda_i=1/2$, 3/2) 
and its vector meson decay amplitude $A^v_{\Lambda_f}$ 
($\Lambda_f=1/2$, 3/2) for
the transverse polarization and $S^v_{\Lambda_f}$ ($\Lambda_f=1/2$) 
for the longitudinal polarization. 
For a resonance of the SU(6)$\otimes$O(3)quark model
with spin $J$, its 12  independent transition amplitudes can be written 
\begin{eqnarray}\label{heli-amp}
H^J_{11} &=& d^J_{1/2,3/2}(\theta) A^v_{1/2} A^\gamma_{3/2},\nonumber\\
H^J_{10} &=& d^J_{-1/2,3/2}(\theta) S^v_{-1/2} A^\gamma_{3/2}\nonumber\\
&=&(-1)^{J_v}d^J_{1/2,3/2}(\pi+\theta) S^v_{1/2} A^\gamma_{3/2},\nonumber\\
H^J_{1-1} &=& d^J_{-3/2,3/2}(\theta) A^v_{-3/2} A^\gamma_{3/2}\nonumber\\
&=&(-1)^{J_v+1}d^J_{3/2,3/2}(\pi+\theta) A^v_{3/2} A^\gamma_{3/2},\nonumber\\
H^J_{21} &=& d^J_{1/2,1/2}(\theta) A^v_{1/2} A^\gamma_{1/2},\nonumber\\
H^J_{20} &=& d^J_{-1/2,1/2}(\theta) S^v_{-1/2} A^\gamma_{1/2}\nonumber\\
&=&(-1)^{J_v}d^J_{1/2,1/2}(\pi+\theta) S^v_{1/2} A^\gamma_{1/2},\nonumber\\
H^J_{2-1} &=& d^J_{-3/2,1/2}(\theta) A^v_{-3/2} A^\gamma_{1/2}\nonumber\\
&=&(-1)^{J_v}d^J_{3/2,1/2}(\pi+\theta) A^v_{3/2} A^\gamma_{1/2},\nonumber\\
H^J_{31} &=& d^J_{3/2,3/2}(\theta) A^v_{3/2} A^\gamma_{3/2},\nonumber\\
H^J_{30} &=& d^J_{1/2,3/2}(\theta) S^v_{1/2} A^\gamma_{3/2},\nonumber\\
H^J_{3-1} &=& d^J_{-1/2,3/2}(\theta) A^v_{-1/2} A^\gamma_{3/2}\nonumber\\
&=&(-1)^{J_v}d^J_{1/2,3/2}(\pi+\theta) A^v_{1/2} A^\gamma_{3/2},\nonumber\\
H^J_{41} &=& d^J_{3/2,1/2}(\theta) A^v_{3/2} A^\gamma_{1/2}\nonumber\\
&=&-d^J_{1/2,3/2}(\theta) A^v_{3/2} A^\gamma_{1/2},\nonumber\\
H^J_{40} &=& d^J_{1/2,1/2}(\theta) S^v_{1/2} A^\gamma_{1/2},\nonumber\\
H^J_{4-1} &=& d^J_{-1/2,1/2}(\theta) A^v_{-1/2} A^\gamma_{1/2}\nonumber\\
&=&(-1)^{J_v}d^J_{1/2,1/2}(\pi+\theta) A^v_{-1/2} A^\gamma_{1/2},
\end{eqnarray}
where in some of the above equations, 
parity conservation allows us to relate $A^v_{-\Lambda_f}$ with 
$A^v_{\Lambda_f}$ for each SU(6)$\otimes$O(3) state with spin $J$ which then 
decays into a vector meson and a nucleon with relative angular momentum 
$J_v$, 
\begin{equation}
A^v_{-\Lambda_f}(J)=(-1)^{1/2-J-J_v}A^v_{\Lambda_f}(J),
\end{equation}
where the factor $1/2$ denotes the spin of the final state nucleon. 
The parity of such a state is $(-1)^N$, where $N$ is the 
main quantum number of the harmonic oscillator shell. 
Meanwhile, the final state system has a parity $(+1)(-1)(-1)^{J_v}$
which is determined by the parity of the nucleon, vector meson, and 
their relative angular momentum. Equivalence of these two parities 
gives $(-1)^N=(-1)^{J_v+1}$, and thus
\begin{equation}\label{symm}
A^v_{-\Lambda_f}(J)=(-1)^{(1/2-J)-(N-1)}A^v_{\Lambda_f}(J).
\end{equation}
So far, the dynamics of the reaction 
has not been explicitly involved. Note that the change of the 
rotation functions has been taken into account in Eq.~(\ref{heli-amp}).
The symmetric feature for each state of SU(6)$\otimes$O(3) quark model
can be seen clearly. 
The nodal structure is determined by the interfering 
amplitudes, therefore can be studied in an explicit model. 

\begin{figure}
\begin{center}
\epsfig{file=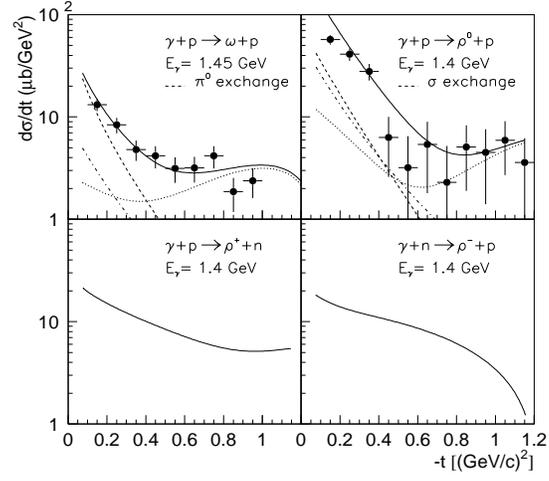,width=8.cm,height=7.cm}
\end{center}
\caption{ Differential cross sections for
four isospin channels. The solid curves denote full-model calculations,
while the dotted and dot-dashed curves denote the exclusive {\it s} 
and {\it u}-channel processes and Pomeron exchanges, respectively.
The dashed curve denotes pion exchange for $\omega$ production, 
and $\sigma$ exchange for $\rho^0$ production, respectively.
Data come from Ref.~\protect\cite{klein}. }
\protect\label{fig:(1)}
\end{figure}

Equations~(\ref{heli-amp}) and ~(\ref{symm})
add some interesting relations to the 12 independent 
transition amplitudes. For example, apart from the rotation function
and phase factors, we can see that 
${\cal H}^J_{1\lambda_v}={\cal H}^J_{3(-\lambda_v)}$, 
and ${\cal H}^J_{2\lambda_v}={\cal H}^J_{4(-\lambda_v)}$,
where ${\cal H}$ denotes the product of $A^v (S^v)$ and $A^\gamma$
in Eq.~(\ref{heli-amp}). 
Thus,
the corresponding amplitudes 
have the same 
dynamical parts. Note that the phase factor and 
the rotation function carry information about the 
resonance partial waves and symmetries, and the dynamical part 
carries information about the resonance structures, which are
generally spin-dependent. 
Therefore, when we add the amplitudes for resonance excitations
together and calculate the cross section differences, 
interferences among resonances 
of different quark model representations will arise.
At low energies, such interferences produce nonzero
contributions to the sum rule. 
We show in Fig.~\ref{fig:(1)} the spin-dependent phenomena near threshold
with the study of cross sections for four reactions.
We shall see below that
at the energy very close to threshold, the 
strong $D_{33}(1700)$ will lead to an 
overestimation of the total cross sections for $\gamma p\to \rho^0 p$.
However, instead of presenting details 
for the study of resonance excitations,
our special attentions will be paid to their 
high energy behavior. 
We expect that at high energies, 
the spin-dependent terms will die out, while spin-independent 
terms will dominate and result in convergence of the cross section
difference.

Certainly, the convergence of the cross section difference
is not obvious in such an approach. 
However, several basic aspects can be outlined. 
Firstly, the quark model wave functions guarantee
the theory to be unitary when all the baryon resonances
are included.
The spatial integrals provide form factors which 
are proportional to $e^{-({\bf k}^2+{\bf q}^2)/6\alpha^2}$
in the transition amplitudes. This factor results in 
the disappearance of resonance contributions at high energies,
which corresponds to high $|{\bf k}|$ and $|{\bf q}|$.
For higher excited states, as shown 
in Ref.~\cite{zhao-prc98}, in the process of photon and meson 
coupling to the same quark, the form factor for the harmonic 
oscillator shell $N$ is
\begin{equation}
F_{Nl}({\bf k}, {\bf q})=\frac{1}{(N-l)!}
\left(\frac{{\bf k}\cdot {\bf q}}{3\alpha^2}\right)^{N-l} 
e^{-({\bf k}^2+{\bf q}^2)/6\alpha^2}, 
\end{equation}
while in the process of photon and meson coupling to different
quarks,
\begin{equation}
F^\prime_{Nl}({\bf k}, {\bf q})=\frac{1}{(N-l)!}(-\frac 12)^N
\left(\frac{{\bf k}\cdot {\bf q}}{3\alpha^2}\right)^{N-l} 
e^{-({\bf k}^2+{\bf q}^2)/6\alpha^2},
\end{equation}
where $N\ge l$ when summed over all permitted $N$ and $l$.
Apparently, an additional factor $(-1/2)^N$
suppresses the process that the photon and meson couple
to different quarks at high energies.
Meanwhile, for a given $N$,
the dominant contribution comes from terms with
$l=0$. 

The dying-out trend
governed by the form factor provides a possible way to bypass 
difficulty arising from resonance excitations. 
A simple analytical argument can be made by  
assuming that those degenerate resonances are on their mass shells
and have the same width. Then, 
the transition amplitudes can be factorized out as
\begin{eqnarray}\label{factorize}
H_{a\lambda_v}
&=&S_{a\lambda_v}({\bf k},{\bf q})
\sum_{N=0}^{\infty}
F_{N0}({\bf k}, {\bf q})
\nonumber\\
&=&S_{a\lambda_v}({\bf k},{\bf q})
e^{-({\bf k}-{\bf q})^2/6\alpha^2},
\end{eqnarray}
where $S_{a\lambda_v}({\bf k},{\bf q})$ is a sum of all 
the spin-dependent and independent parts.
Note that, we neglect the process of photon and meson coupling to different
quarks, and take the leading contribution of $l=0$.
The second line of Eq.~(\ref{factorize}) could be regarded as a 
good approximation for the resonance excitations at $\nu\to \infty$.
The behavior of the exponent, despite its angular-dependence, 
suppresses the amplitudes
with increasing energies, especially at backward direction.
This feature is very helpful since those spin-dependent
terms generally have large effects at large angles.
Remember that 
$S_{a\lambda_v}({\bf k},{\bf q})$ contains both spin-dependent and 
independent terms, 
a simple survey over these amplitudes 
indeed suggests that the spin-dependent terms are
strongly suppressed at high energies.
It can be seen quite explicitly that the spin-independent
term $(\veps_v\times{\bf q})\cdot (\veps_\gamma\times{\bf k})$ 
($=\veps_v\cdot\veps_\gamma {\bf q}\cdot{\bf k}-\veps_v\cdot 
{\bf k}\veps_\gamma\cdot{\bf q}$)
plays a dominant role, 
where $\veps_v$ is the transverse polarization vector 
for the vector meson. 
This feature is the first aspect that guarantees the convergence
of the integral at a few GeV's.

The spin-dependence of the Pomeron exchange 
in the cross section difference calculations
is also not obvious due to interferences among different
spin operators. Fortunately, however, some typical
features of the Pomeron exchange 
model~\cite{donnachie,laget95,lee96-97} can help us 
understand its behavior at high energies.
Since the Pomeron, which accounts for the 
diffractive process, is rather like a charge conjugation
$C=+1$ vector meson,
this feature means that its dominant contribution
is located at small $|t|$. 
The longitudinal amplitude becomes negligible 
at high energies. As discussed in detail in Ref.~\cite{zhao-npa99},
the dominant term is proportional to $\veps_\gamma\cdot\veps_v$ in the 
c.m. system, which is spin-independent. 
When $\nu\to\infty$,
we can immediately see that its contribution to the cross section
difference $\Delta\sigma(\nu)$ is zero, i.e., 
$H^2_{1\lambda_v}+H^2_{3\lambda_v}=H^2_{2\lambda_v}+H^2_{4\lambda_v}$.
In other words, the Pomeron exchange contribution to the GDH
sum rule becomes negligible at high energies, although it is
dominant over all other processes.

Explicitly, the equivalence,
$H^2_{1\lambda_v}+H^2_{3\lambda_v}=H^2_{2\lambda_v}+H^2_{4\lambda_v}$,
can be satisfied in the exclusive $\pi^0$ (for $\omega$ and $\phi$) 
and $\sigma$ exchange (for $\rho^0$) due to the feature that
no spin carried by the exchanged pion and $\sigma$ meson.
Undoubtedly, the interferences from the resonance excitations at low energies
will violate the equivalence.  
However,
beyond the resonance region, all these contributing processes
will be dominated by the spin-independent terms, which 
will result in vanishing of $\Delta\sigma(\nu)$.
Numerically, we find that 
when $\nu\approx 6$ GeV, $\Delta\sigma(\nu)$ becomes negligible. 
Above the energy, the effects from 
spin-dependent terms will only make a change to the fourth decimal place
in the integrals.
This gives confidence  
of cutting off the photon energies
at a few GeV's for our model.

\begin{figure}
\begin{center}
\epsfig{file=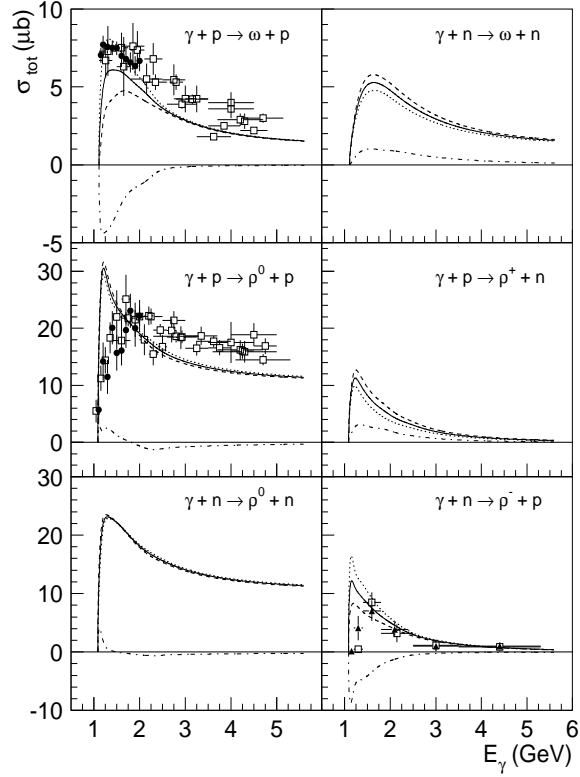,width=10.cm,height=14.cm}
\end{center}
\caption{ Total cross sections and cross section differences for 
the $\omega$ and $\rho$ meson photoproduction. 
The solid, dashed, dotted, and dot-dashed curves
denote the calculations for $\sigma_{tot}$ 
[$=(\sigma_{1/2}+\sigma_{3/2})/2$], $\sigma_{1/2}$,
$\sigma_{3/2}$, and $\Delta\sigma$
($=\sigma_{1/2}-\sigma_{3/2}$), respectively.
Data come from Refs.~\protect\cite{klein,old-data,benz74}. }
\protect\label{fig:(2)}
\end{figure}

In Fig.~\ref{fig:(2)}, we 
report the calculations of the cross section differences
of six isospin channels. 
The parameters are extracted from the $\omega$ production channel.
We do not attempt to fit data for the $\rho$ productions.
This requires more subtle considerations. 
In fact, the prediction gives an overall agreement
with the data, which suggests that isospin 
conservation has been roughly kept for the $\omega$ and $\rho$
mesons.
A sensible feature arising between the $\omega$ and
$\rho$ production is that isospin 3/2 resonances
will contribute
in the $\rho$ meson production but be eliminated in the $\omega$
production,
as required by isospin conservation.
Meanwhile, the quark model symmetry 
eliminates those states of quark model 
representation $[{\bf 70}, ^{\bf 4} {\bf 8}]$
from contribution in the proton target reactions~\cite{moorhouse}.
This feature results in an interchanging of the relative positions
between the dashed ($\sigma_{1/2}$) and dotted curves ($\sigma_{3/2}$)
in $\gamma p\to \omega p$ and $\gamma n\to \omega n$. 
Similar phenomena can be also seen in $\gamma p\to \rho^+ n$ and 
$\gamma n\to \rho^- p$.
It is still worth noting that the $D_{33}(1700)$
is found to play an important role in the $\rho$ meson 
production near threshold. Details of this will be reported 
and discussed elsewhere.

The contribution of the $\omega$ and $\rho$ channels 
to the GDH sum rule are listed in Table~\ref{tab:(1)}
to compare with other channel contributions predicted
by other studies. 
The neutral $\rho$ production has a relatively
smaller contribution to the proton and neutron sum rule,
while contributions from the $\omega$ and $\rho^\pm$ channels
are more than one order of magnitude larger. 
However, when we sum over those reaction channels for the proton and 
neutron, we find that the overall magnitude of 
the contribution to the proton ($I^v_{GDH}=+0.26 \ \mu$b) is 
much smaller than to the neutron ($I^v_{GDH}=-2.05 \ \mu$b).
In particular, the sign of the sum ($I^v_{GDH}$)
suggests that the contribution to the proton sum rule will cancel 
a small number of previous results~\cite{tiator},
while the contribution
to the neutron will add a relatively larger one. 
This trend is fairly consistent with the exclusive 
studies of Ref.~\cite{tiator}.

In conclusion, we have evaluated the contributions 
of vector meson photoproduction to the GDH sum rule 
using a quark model with an effective Lagrangian. 
Although more detailed study at resonance region is needed,
we show that
the cross section differences converge
with the increasing energy, thus, leads to a reasonable 
energy cut at 6 GeV. 
The main contributions 
of vector meson photoproduction
are found from the resonance region.
Although the total values are small,
their corrections to the sum rule are shown to be 
in the right direction.
This study provides a test for our approach for vector meson 
photoproduction though more accurate data from experiments 
are needed to constrain the model.
Nevertheless, it  
provides some insights into the relevance of the exclusive 
vector meson photoproduction to the GDH sum rule.

Stimulating discussions with F.E. Close and Z.-P. Li
are gratefully acknowledged. 
We would like to thank L. Tiator for 
very helpful comments.
The financial support of the U.K. Engineering and Physical 
Sciences Research Council (Grant No. GR/M82141) is gratefully 
acknowledged.

%\acknowledgements

%%
\begin{table}
\caption{ Contributions of vector meson photoproduction to the GDH sum rule
in comparison with other exclusive channels. Contributions from $\pi N$,
$\eta N$, $\pi\pi B$ (Born terms), and $\pi\pi D$ [$D_{13}(1520)$ resonance]
are from Mainz group study~\protect\cite{tiator}; kaon channel contributions
are from Ref.~\protect\cite{sumowidagdo}; 
experimental data are from Ref.~\protect\cite{gdh-a2}.}
\protect\label{tab:(1)}
\begin{center}
\begin{tabular}{lc|lc}
%\hline\hline\\[1ex]
Proton & $I_{GDH}$ ($\mu$b) 
& Neutron & $I_{GDH}$ ($\mu$ b) \\[1ex]\hline
$\gamma p\to \pi^0 p$ & $-150$ & $\gamma n\to \pi^0 n$ & $-154$ \\[1ex]
$\gamma p\to \pi^+ n$ & $-21$ & $\gamma n\to \pi^- p$ & $+30$ \\[1ex]
$\gamma p\to \eta p$ & $+15$ & $\gamma n\to \eta n$ & $+10$ \\[1ex]
$\gamma p\to \pi\pi B$ & $-30$ & $\gamma n\to \pi\pi B$ & $-35$ \\[1ex]
$\gamma p\to \pi\pi D$ & $-15$ & $\gamma n\to \pi\pi D$ & $-15$ \\[1ex]
$\gamma p\to K^+\Lambda$ & $+1.66$ & $\gamma n\to K^0\Lambda$ & $-4.78$\\[1ex]
$\gamma p\to K^+\Sigma^0$ & $+1.53$ & $\gamma n\to K^+\Sigma^-$& $+1.59$\\[1ex]
$\gamma p\to K^0\Sigma^+$ & $+0.83$ & $\gamma n\to K^0\Sigma^0$& $+1.21$\\[1ex]
$\gamma p\to \omega p$ & $-2.01$ & $\gamma n\to \omega n$ & $+0.93$ \\[1ex]
$\gamma p\to \rho^0 p$ & $+0.05$ & $\gamma n\to \rho^0 n$ & $-0.05$ \\[1ex]
$\gamma p\to \rho^+ n$ & $+2.22$ & $\gamma n\to \rho^- p$ & $-2.93$ \\[1ex]\hline
sum of above & $-196.72$ & sum of above & $-168.03$\\[1ex]
GDH & $-205$ & GDH & $-233$ \\[1ex]
exp. results & $-210$~\footnote{In Ref.~\cite{gdh-a2}, 
$I_{GDH}=-226\pm$5(stat)$\pm$12 (syst) $\mu$b was reported in the energy range
$200<\nu <800$ MeV, while taking into account the missing contributions 
from $\nu <200$ MeV, a deduced value $-210 \ \mu$b was estimated.} 
& exp. results & not available \\[1ex]
\end{tabular}
\end{center}
\end{table}

\end{document}